\documentclass[a4paper,twoside]{article}

\usepackage{epsfig}
\usepackage{subcaption}
\usepackage{calc}
\usepackage{amssymb}
\usepackage{amstext}
\usepackage{amsmath}
\usepackage{amsthm}
\usepackage{multicol}
\usepackage{pslatex}
\usepackage{apalike}
\usepackage{todonotes}
\usepackage{SCITEPRESS}     

\begin{document}

\title{Combining Gesture and Voice Control for Mid-Air Manipulation of CAD Models in VR Environments}

\author{\authorname{Markus Friedrich\sup{1}, Stefan Langer\sup{1}and Fabian Frey\sup{1}}
\affiliation{\sup{1}Institute of Informatics, LMU Munich, Oettingenstraße 67, 80538 Munich, Germany}
\email{\{markus.friedrich, stefan.langer\}@ifi.lmu.de, fabian.frey@campus.lmu.de}}


\abstract{
Modeling 3D objects in domains like Computer Aided Design (CAD) is time-consuming and comes with a steep learning curve needed to master the design process as well as tool complexities.
In order to simplify the modeling process, we designed and implemented a prototypical system that leverages the strengths of Virtual Reality (VR) hand gesture recognition in combination with the expressiveness of a voice-based interface for the task of 3D modeling.
Furthermore, we use the Constructive Solid Geometry (CSG) tree representation for 3D models within the VR environment to let the user manipulate objects from the ground up, giving an intuitive understanding of how the underlying basic shapes connect.
The system uses standard mid-air 3D object manipulation techniques and adds a set of voice commands to help mitigate the deficiencies of current hand gesture recognition techniques. 
A user study was conducted to evaluate the proposed  prototype.
The combination of our hybrid input paradigm shows to be a promising step towards easier to use CAD modeling.
}
\onecolumn \maketitle \normalsize \setcounter{footnote}{0} \vfill

\section{\uppercase{Introduction}}
\label{sec:introduction}
Current Computer Aided Design (CAD) modeling tools use mainly mouse control-based object manipulation techniques in predefined views of the modelled objects.
However, manipulating 3D objects with a 2D input device (mouse) on a 2D output device (monitor) always necessitates a complex transfer, making it difficult for beginners to grasp the needed interaction concepts.
\\
With the advent of affordable Virtual Reality (VR) devices and robust hand gesture recognition systems, more possibilities are at the fingertips of interaction designers and researchers.
The goal is to leverage the potential of these immersive input methodologies to improve intuitiveness, flattening the learning curve, and increasing efficiency in the 3D modeling task. 
\\
However, hand gesture recognition systems face many challenges in productive environments.
Compared to expert devices like 3D mouses, these systems lack robustness as well as precision. 
Moreover, constant arm movements can cause fatigue syndroms like the Gorilla Arm Syndrom \cite{lavalle2017}, making it hard to use systems over an extended period of time. 
This is where we hypothesize that a hybrid approach, combining gesture recognition and voice control, is beneficial.
While using hands for intuitive model part manipulation, voice commands replace complex to recognize gestures with simple to say and easy to memorize word commands.
\\
In this work, we propose a novel interaction concept, implemented as a prototype, which combines hand gesture recognition and voice control for CAD model manipulation in a VR environment. 
The prototype uses gesture-based mid-air 3D object manipulation techniques where feasible and combines it with a set of voice commands to complement the interaction concept.
CAD models are represented as a combination of geometric primitives and Boolean set-operations (so called Constructive Solid Geometry (CSG) trees) enabling transformation operations that are more intuitive for beginners.
The whole system is evaluated in a user study, showing its potential. 
\\
The paper makes the following contributions: 
\begin{itemize}
    \item A new interaction concept for intuitive, CSG tree-based, CAD modeling in VR leveraging the strengths of both, gesture- and voice-based interactions.
    \item A prototypical implementation of the interaction concept with off-the-shelf hard- and software.
    \item A detailed user study proving the advantages of the proposed approach.
\end{itemize}
The paper is structured as follows:
The problem solved by the proposed system is detailed in Section \ref{sec:probdesc}.
Essential terms are explained in Section \ref{sec:background}. 
Related work is discussed in Section \ref{sec:related_work} which focuses on mid-air manipulation techniques. 
This is followed by an explanation of the concept (Section \ref{sec:concept}) which is evaluated in Section \ref{sec:evaluation}. 
Since the proposed prototype opens up a multitude of different new implementation and research directions, the paper concludes with a summary and a thorough description of possible future work (Section \ref{sec:futurework}).

\section{\uppercase{Problem Description}}
\label{sec:probdesc}
The goal is to provide a system with which the user can manipulate 3D CAD models in an intuitive way, leveraging gesture and voice recognition techniques in a VR environment. 
CAD models are represented as CSG tree structures which combine geometric primitives (spheres, cylinders, boxes) with Boolean set-operations (intersection, union, difference).
This representation was chosen since it is proven to be highly intuitive, supporting our goal of designing an easy-to-learn modeling system and interaction concept.
This also implies that the user should be able to modify both, geometric primitives with basic transformations (scaling, rotating, translating), as well as properties of the underlying CSG tree.

\section{\uppercase{Background}}
\label{sec:background}
\subsection{Construction Trees}
Construction or CSG trees \cite{requicha80} is a representation and modeling technique for geometric objects mostly used in CAD use cases. 
Complex 3D models are created by combining primitive geometric shapes (spheres, cylinders, convex polytopes) with Boolean operators (union, intersection, difference, complement) in a tree-like structure.  
The inner nodes describe the Boolean operators while the leave nodes represent primitives. 
The CSG tree representation has two big advantages over other 3D model representations: Firstly, it is memory-saving, and secondly, it is intuitive to use. 

\subsection{Mid-Air Interactions for 3D Object Manipulation}
The manipulation of 3D objects in a virtual scene is a fundamental interaction in immersive virtual environments. 
Manipulation is the task of changing characteristics of a selected object using spatial transformations \cite{bowman1999formalizing}.  
Translation, rotation, and scaling are referred to as basic manipulation tasks \cite{bowman20043d}. 
Each task can be performed in any axis direction. 
In general, a single transformation in one specific axis is defined as a degree of freedom (DOF). 
Thus, a system that provides a translation-only interface on all three axes supports three DOFs whereas a system that offers all three manipulation tasks on all axes has 9 DOFs \cite{mendes2019survey}.
\\
In so-called mid-air interactions, inputs are passed on through the user's body, including posture or hand gestures \cite{mendes2019survey}. 
This kind of interaction provides a particularly immersive and natural way to grab, move, or rotate objects in VR which allows for more natural interfaces that can increase both, usability and user performance \cite{caputo2019gestural}.
In this paper, the term mid-air manipulation refers specifically to the application of basic transformations to virtual objects in a 3D environment using hand gestures. 
Apart from the allowed degrees of freedom, existing techniques can be classified by the existence of a separation between translation, rotation, and scaling \cite{mendes2019survey} which also affects manipulation precision.
Furthermore, two additional interaction categories exist: Bimanual and unimanual. 
Bimanual interfaces imply that users need both hands to perform the intended manipulation, whereas applications using unimanual interfaces can be controlled with a single hand \cite{mendes2019survey}.
\\
Helpful guidelines for the design of mid-air object manipulation interaction concepts can be found in \cite{mendes2019survey} which were also considered during the design of the proposed system.

\section{\uppercase{Related Work}}
\label{sec:related_work} 
\begin{figure}[h!]
	\centering
	\subfloat[Handle Box] {{\includegraphics[width=0.45\linewidth]{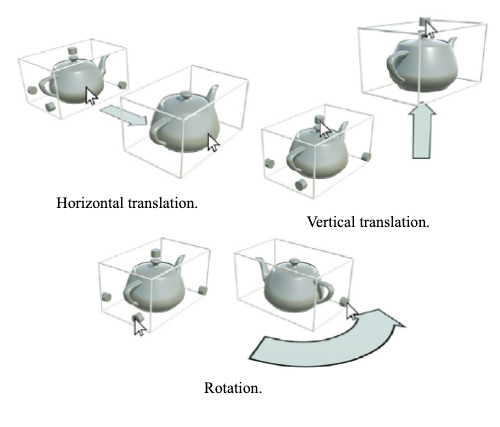}}}%
	\qquad
	\subfloat[Virtual Handles]
	{\includegraphics[width=0.45\linewidth]{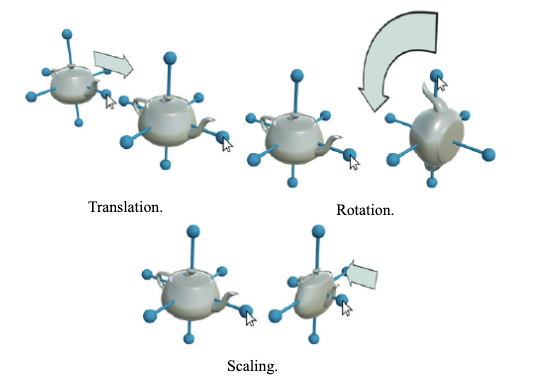}}%
	\caption{Mouse manipulation techniques. Source: \cite{mendes2019survey}.}%
	\label{fig:handles}%
\end{figure}

\begin{figure}[h!]
	\centering
	\includegraphics[width=0.7\linewidth]{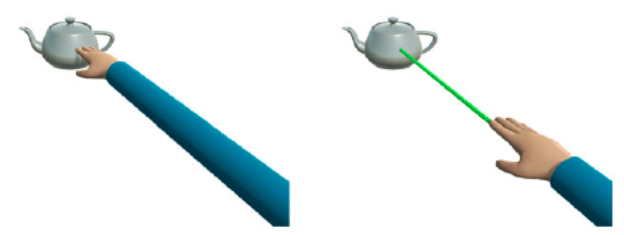}
	\caption{Go-Go (left) and Ray-casting (right) technique. Source: \cite{mendes2019survey}.}
	\label{fig:gogo}
\end{figure}

\begin{figure}[h!]
	\centering
	\includegraphics[width=0.7\linewidth]{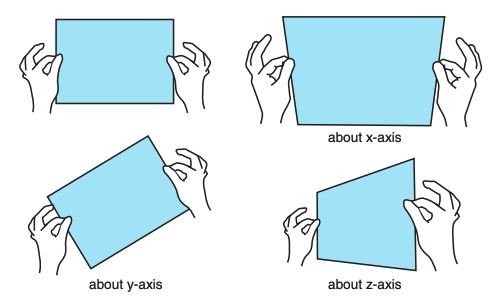}
	\caption{Manipulation based on the sheet of paper metaphor. Source: \cite{wang20116d}.}
	\label{fig:sheet}
\end{figure}

\begin{figure}[h!]
	\centering
	\includegraphics[width=0.9\linewidth]{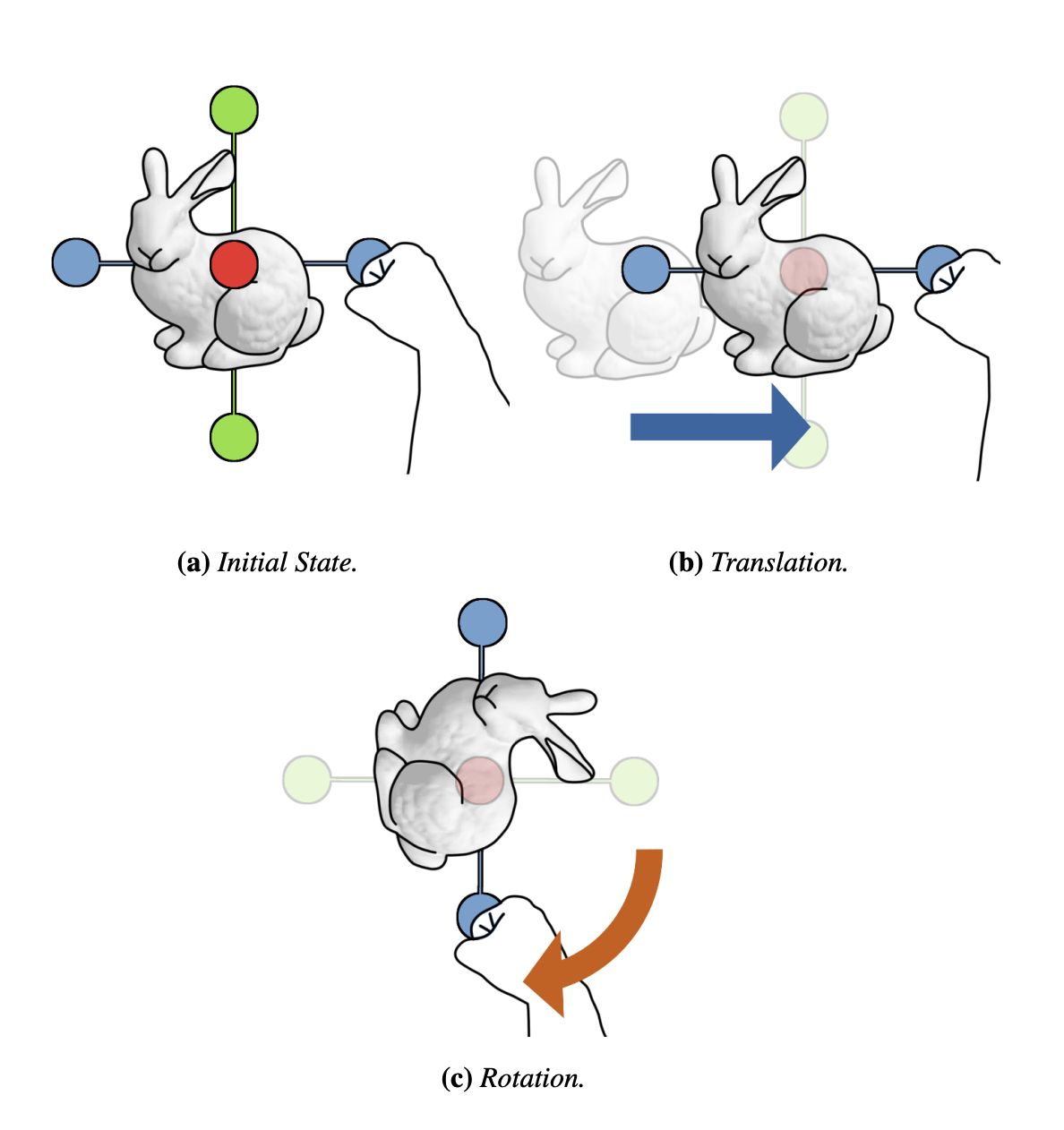}
	\caption{Virtual Handles in VR. Source: \cite{mendes2016benefits}.}
	\label{fig:virtualhandle}
\end{figure}

This section provides an overview of interaction concepts for CAD modelling in VR environments. A detailed survey on this topic is provided in \cite{mendes2019survey}.
\\
\textbf{Object Manipulation.}
The so-called Handle Box approach \cite{houde1992iterative} is essentially a bounding box around a selected object. 
The Handle Box consists of a lifting handle, which moves the object up and down, as well as four turning handles to rotate the object around its central axis. 
To move the object horizontally, an activation handle is missing and instead, the user can easily click and slide (see Figure \ref{fig:handles} (left)).
\\
In \cite{conner1992three}, so-called Virtual Handles that allow full 9 DOF control are proposed. 
The handles have a small sphere at their ends, which are used to constrain geometric transformations to a single plane or axis \cite{mendes2019survey}. 
The user selects the manipulation mode with the mouse button. 
During rotation, the initial user gesture is recognized to determine the rotation axis (see Figure \ref{fig:handles} (right)).
Both techniques are initially designed for classic mouse/screen interactions but are seamlessly transferable to hand gesture/VR environments. 
\\
\textbf{Mid-Air Interactions.}
Caputo et al. state that users often like mid-air interactions more than other methods, finding the accuracy in manipulation sufficiently good for many proposed tasks \cite{caputo2019gestural}. 
First systems used gloves or other manual input devices providing the functionality to manipulate or to grab objects. 
In \cite{robinett1992implementation}, a system for translating, rotating, and scaling objects in VR is proposed.
In \cite{bowman1997evaluation}, the Go-Go and the Ray-casting technique are proposed which allow users to grab objects which are farther away than an arm's length (see Figure \ref{fig:gogo}). 
For now, our concept focuses on small objects, leaving this aspect for future work.
The formalization of the object manipulation task in three sub tasks, namely (object) selection, (object) transformation and (object) release was proposed in \cite{bowman1999formalizing}.
One of the most influential techniques in this field is the Hand-centered Object Manipulation Extending Ray-casting (HOMER) method  \cite{bowman1997evaluation}.
After selecting an object with a light ray, the hand moves to the object. 
The hand and the object are linked and the user can manipulate the object with a virtual hand until the object has been dropped. 
After that, the connection is removed and the hand returns to its natural position.
\\
\textbf{Metaphors For Manipulation.}
In the so-called Spindle system \cite{mapes1995two}, the center point of both hands is used to select objects. 
Transformation is done by moving both hands simultaneously. 
Hand movements around the center point rotate the object and changing the distance between both hands scales it.
\\
A bimanual 7 DOF manipulation technique is the Handlebar introduced by Song et al. \cite{song2012handle}. 
It is adopted by multiple real-world applications. 
The Handlebar uses the relative position and movement of both hands to identify translation, rotation, scaling, and to map the transformation into a virtual object. 
It was developed for low-cost hardware with reasonable accuracy without any need for precise finger detection. 
The main strength of this method is the intuitiveness due to the physical familiarity with real-world actions like rotating and stretching a bar. 
However, holding an arm position may exhaust the user (Gorilla Arm Syndrome \cite{lavalle2017}). 
\\
In \cite{wang20116d}, manipulations (in particular translation and rotation) are separated.
Translation is applied by moving the closed hand and rotation by using two hands around the three main axes, mimicking the physical action of rotating a sheet of paper (see Figure \ref{fig:sheet}).
\\
Another bimanual metaphor is the so-called 2HR technique proposed by Caputo et. al \cite{caputo2015evaluation}, which assigns interaction manipulation to both hands.
Movements of the dominant hand trigger a translation of objects whereas the non-dominant hand's rotation is mapped to the object rotation. 
However, overall performance of this technique is poor in contrast to the Handlebar metaphor \cite{caputo2015evaluation}. 
\\
The most direct unimanual method is the Simple Virtual Hand metaphor \cite{caputo2019gestural} where objects are selected by grasping them. 
After the selection, movement and rotation of the user's hand are mapped directly to the objects. 
This method is characterized by its high intuitiveness.
However, accurate, reliable and robust hand tracking is required. 
Kim et al. extended this idea in \cite{kim20143d} by placing a sphere around the selected object and additionally enable the user to control the reference frame during manipulation which shows great improvements, in particular for rotation.
\\
In order to demonstrate the advantages of DOF separation, Mendes et al. applied the Virtual Handles concept \cite{conner1992three} to VR environments \cite{mendes2016benefits} as depicted in Figure \ref{fig:virtualhandle}. 
Since Virtual Handles allow users to choose a single axis, transformations in unwanted axes are improbable.
However, transformations in more than one axis take a bit more time.
We use this concept for basic transformations in our prototype. 
\\
In the field of mid-air interactions open challenges still exist. 
For example, there are no widely accepted manipulation metaphors \cite{caputo2019gestural} and most techniques follow a direct mapping strategy with multiple DOFs controlled at the same time, which is only suitable for coarse transformations \cite{mendes2019survey}. 
 
\section{\uppercase{Concept}}
\label{sec:concept}

In order to simplify the modeling process as much as possible, we utilized mid-air manipulation techniques without any controllers or gloves - hand tracking only. 
The main feature of our technique is the use of only the grasping gesture. 
This simple gesture allows the usage of low-cost hand-tracking sensors, more specifically the Leap Motion Controller sold by Ultrahaptics.
More complex instructions are operated via voice control.
To the best of our knowledge, this is the first approach that combines mid-air manipulation techniques with voice commands in order to create an intuitive and fast way of manipulating 3D objects in VR.
The design of the system is presented in the following sections. 

\subsection{System Overview}
\begin{figure}
	\centering
	\includegraphics[width=1.0\linewidth]{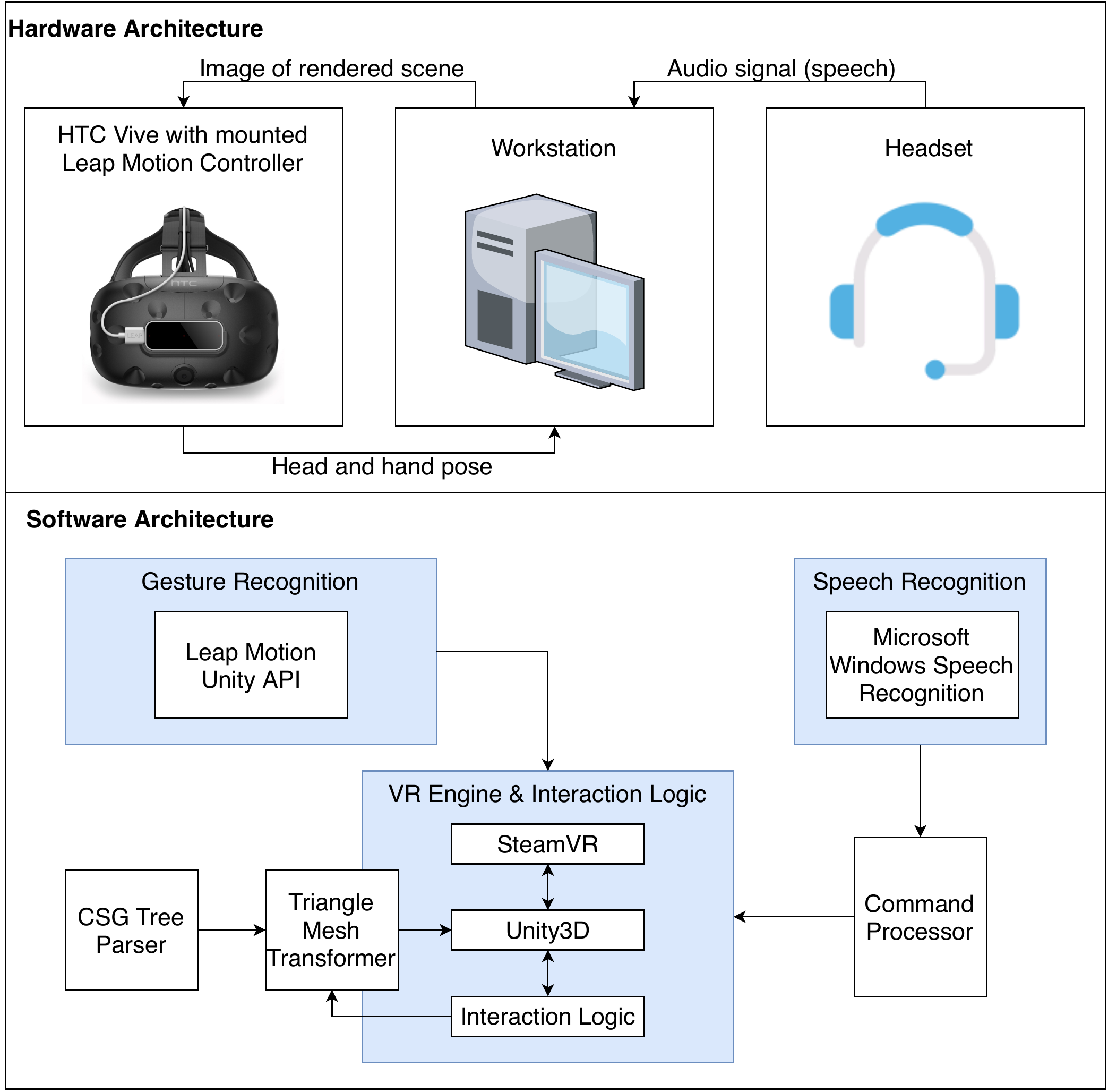}
	\caption{Architecture overview of the developed system.}
	\label{fig:systemoverview}
\end{figure}

The proposed system consists of hardware and software components and is detailed in Figure \ref{fig:systemoverview}.
The hand tracking controller (\textit{Leap Motion Controller}) is mounted on the VR headset which is an off-the-shelf \textit{HTC Vive} that uses laser-based head tracking and comes with two $1080$x$1200$ $90$Hz displays.
The speech audio signal is recorded using a wearable microphone (\textit{Headset}). 
A central computing device (\textit{Workstation}) takes the input sensor data, processes it, applies the interaction logic and renders the next frame which is then displayed in the VR device's screens. 
\\
The software architecture consists of several sub-components. 
The \textit{Gesture Recognition} module translates input signals from the \textit{Leap Motion sensor} into precise hand poses and furthermore recognizes hand gestures by assembling temporal chains of poses. The \textit{Speech Recognition} module takes the recorded speech audio signal and translates it into a textual representation. 
The CSG model, which is later used for editing, is parsed by the \textit{CSG Tree Parser} based on a JSON-based input format and transformed into a triangle mesh by the \textit{Triangle Mesh Transformer}. This step is necessary since the used rendering engine (\textit{Unity3D}) is restricted to this particular geometry representation.
The main component (\textit{VR Engine \& Interaction Logic}) reacts on recognized gestures and voice commands, executes corresponding actions (\textit{Interaction Logic}), and handles VR scene rendering (\textit{Unity3D}, \textit{SteamVR}).
For example, if the user applies a basic scaling transformation to a model part consisting of a sphere, the geometric parameters of the sphere (in this case its radius) are changed based on the hand pose.
Then, the model is re-transformed into the triangle mesh representation, rendered, and the resulting image is sent to the VR device.

\subsection{Interaction Concept}
The interaction concept is based on hand gestures (visible as two virtual gloves that 
visualize the user's hand and finger poses) and simple voice commands.
It furthermore consists of two tools, the model tool and the tree tool, as well as the information board serving as a user guidance system. 

\subsubsection{Model Tool}
\begin{figure}[h!]
	\centering
	\includegraphics[width=0.7\linewidth]{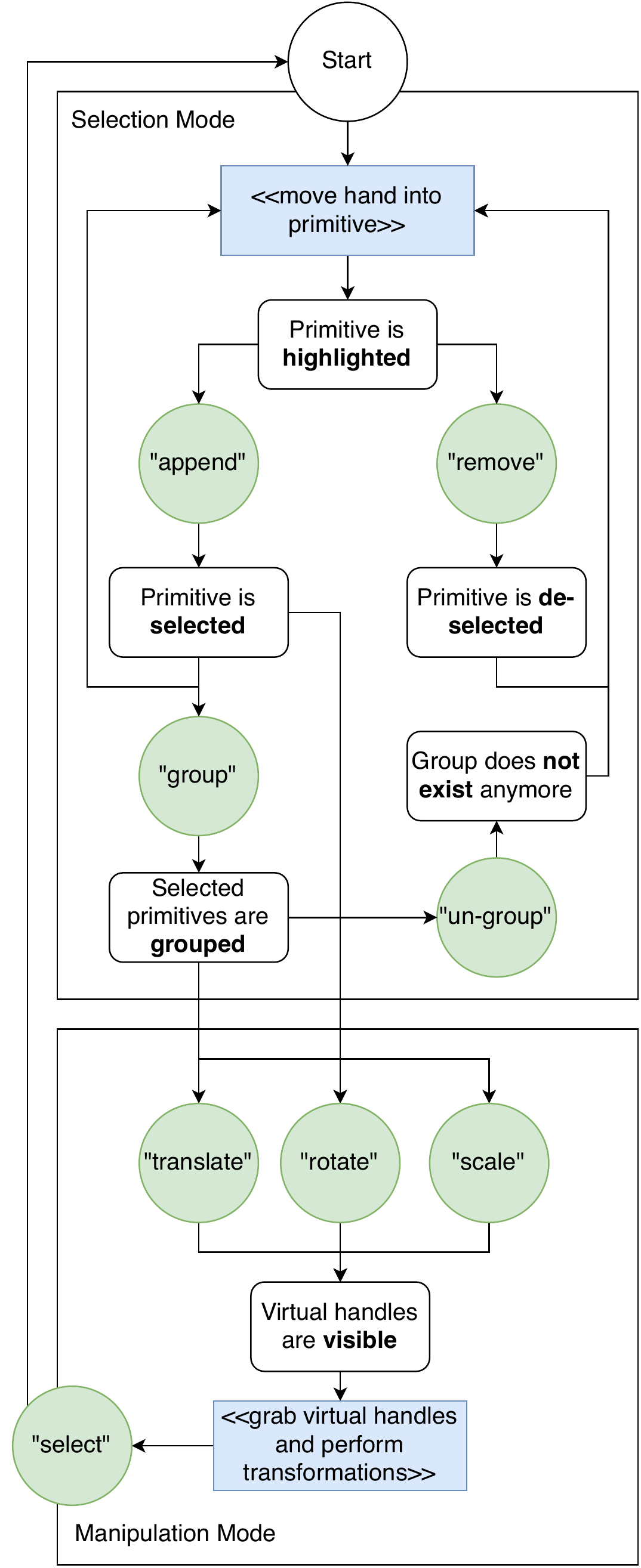}
	\caption{Interaction concept of the model tool.}
	\label{fig:mt_interaction}
\end{figure}

\begin{figure}[h!]
 	\centering
 	\subfloat[Initial State.]{\includegraphics[width=0.45\linewidth]{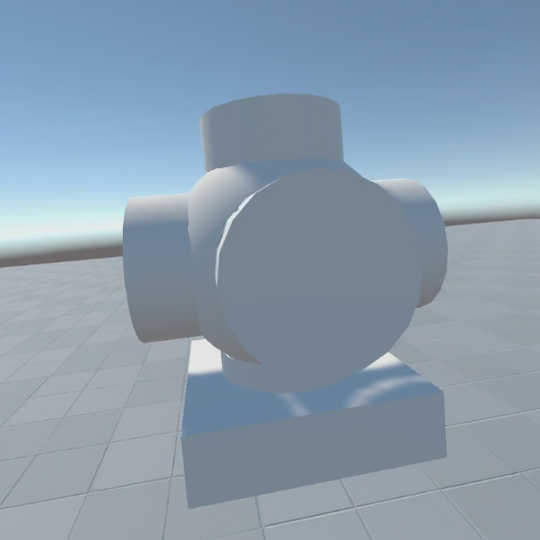}}%
 	\qquad
 	\subfloat[Highlighted.]{\includegraphics[width=0.45\linewidth]{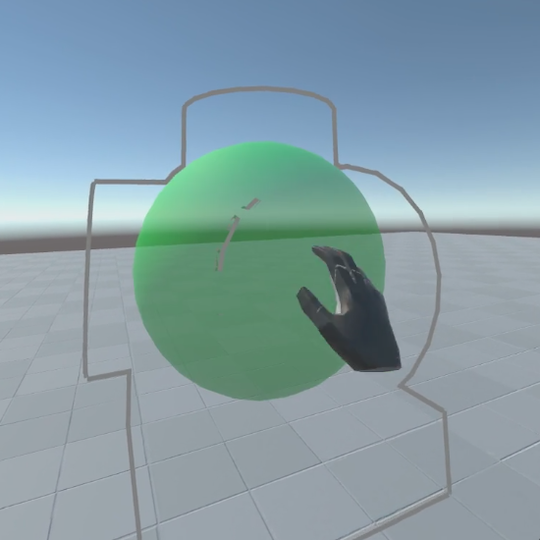}}
    \par\bigskip
    \subfloat[Selected.]{\includegraphics[width=0.45\linewidth]{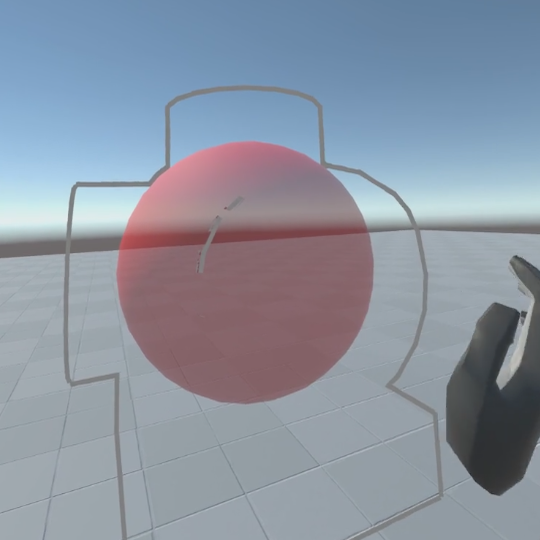}}
 	\qquad
 	\subfloat[Grouped.]{\includegraphics[width=0.45\linewidth]{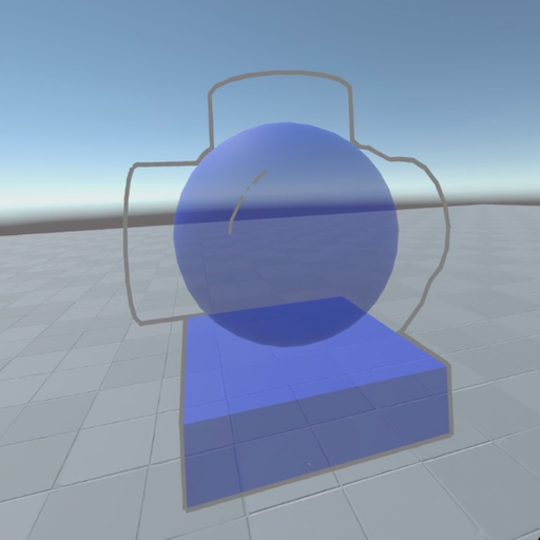}}
 	\caption{Illustration of all possible states of a primitive.}%
 	\label{fig:states}%
 \end{figure}

\begin{figure}[h!]
	\centering
	\includegraphics[width=0.43\linewidth]{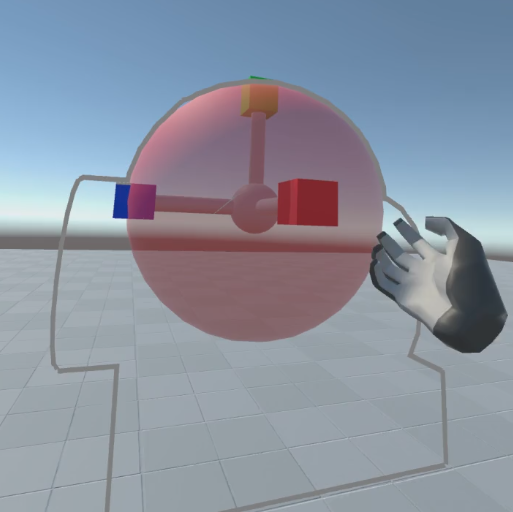}
	\caption{Virtual Handles are used for all basic transformations in manipulation mode.}
	\label{fig:virtual_handle}
\end{figure}

The model tool is used for direct model manipulation and is active right after system startup.
It can be divided in two different input modes, the selection mode and the manipulation mode. 
Modes can be switched using the voice command 'select' (to selection mode) and one of the manipulation commands 'scale', 'translate' and 'rotate' (to manipulation mode). 
Figure \ref{fig:mt_interaction} details the interaction concept of the model tool.
\\
The main reason for separating interactions that way is the ambiguity of hand gestures. 
In some cases, the recognition system was unable to separate between a model part selection and a translation or rotation transformation since both use the same grasp gesture.
The grasp gesture was chosen since initial tests which involved basic recognition tasks for all supported gestures revealed that it is the most robustly recognized gesture available. 
\\
\textbf{Selection Mode.}
The initial application state can be seen in Figure \ref{fig:states} (a): 
The triangle mesh of the loaded model is displayed in grey. 
The user can enter the selection mode by saying the word 'select'. 
The mode change is additionally highlighted through an on-screen message.
Using virtual hands, the user can enter the volume of the model which highlights the hovered primitives in green (Figure \ref{fig:states} (b)). 
This imitates the hovering gesture well known from desktop-based mouse interfaces.
Once highlighted, the user can append the primitive to the list of selected primitives by using the voice command 'append'.
Selected primitives are rendered in red (Figure \ref{fig:states} (c)). 
This way, multiple primitives can be selected. 
In order to remove a primitive from the selection, the user's hand must enter the primitives volume and use the voice command 'remove'.
Multiple selected primitives can be grouped together with the voice command 'group' which is useful in situations where the selected primitives should behave like a single primitive during manipulation, e.g., when a rotation is applied.
A group is displayed in blue (see Figure \ref{fig:states} (d)) and can be dissolved by saying 'un-group'.
\\
\textbf{Manipulation Mode.}
For the manipulation of selected or grouped primitives (in the following: a model part), three basic transformations are available: translation, rotation and scaling.
The currently used transformation is selected via a voice command ('translate', 'rotate' and 'scale').
The manipulation mode is entered automatically after invoking one of these commands. 
The selected transformation is highlighted through the displayed Virtual Handles \cite{conner1992three} (three coordinate axes with a small box at their ends, pointing in x-, y-, and z-direction as shown in Figure \ref{fig:virtual_handle}).
\\
\textit{Rotation}. 
Per-axis rotations are done by grabbing the small boxes at the end of each coordinate axis and performing a wrist rotation.
The grab gesture is depicted in Figure \ref{fig:gestures} (a).
The rotation is directly applied to the corresponding model part axis. 
Alternatively, the sphere displayed at the center of the coordinate system can be grabbed and rotated. 
The sphere rotations are directly applied to the model part. 
This by-passes the restrictions of per-axis rotations and allows for faster manipulation at the cost of precision. 
\\
\textit{Translation}. 
Per-axis translations work like per-axis rotations. 
However, instead of rotating the wrist, the grabbed boxes can be moved along the corresponding axis which results in a direct translation of the model part. The sphere in the center of the coordinate axis can be grabbed and moved around without any restrictions. 
Resulting translations are directly applied to the model part. 
\\
\textit{Scaling}. 
Per-axis scaling works similar to per-axis translation. 
The use of the sphere at the center of the coordinate system is not supported since it cannot be combined with a meaningful gesture for scaling.
\\
The transformation to use can be switched in manipulation mode by invoking the aforementioned voice commands. 
If the primitive selection or group should be changed, a switch to the selection mode is necessary.

\subsubsection{Tree Tool}
\begin{figure}[h!]
	\centering
	\includegraphics[width=0.43\linewidth]{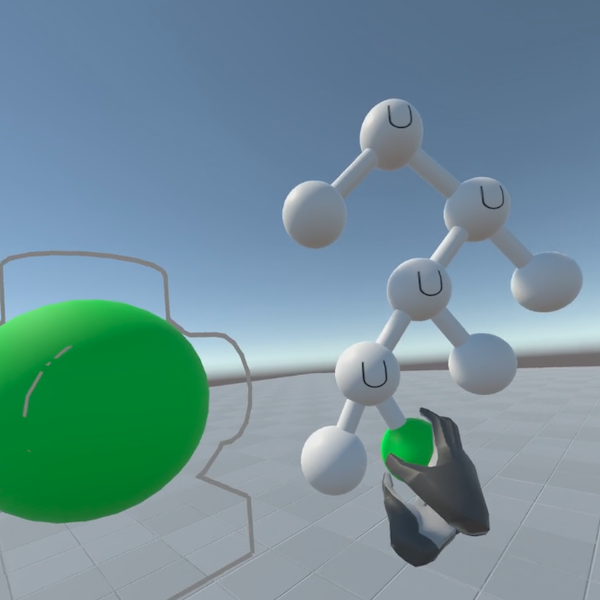}
	\caption{Highlighting a primitive using the tree tool.}
	\label{fig:tree}
\end{figure}

\begin{figure}[h!]
	\centering
	\subfloat[While the hand is not pointing upwards, the tree is hidden.]{\includegraphics[width=0.4\linewidth]{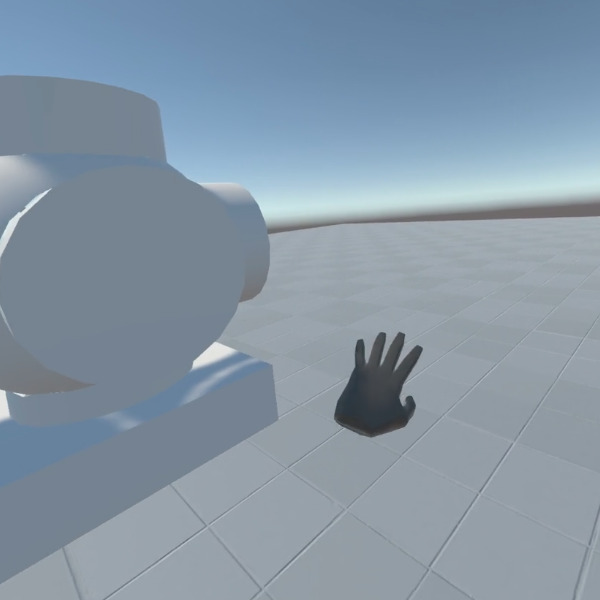}}%
	\qquad
	\subfloat[The tree is displayed when the palm of the hand points upwards.]{\includegraphics[width=0.4\linewidth]{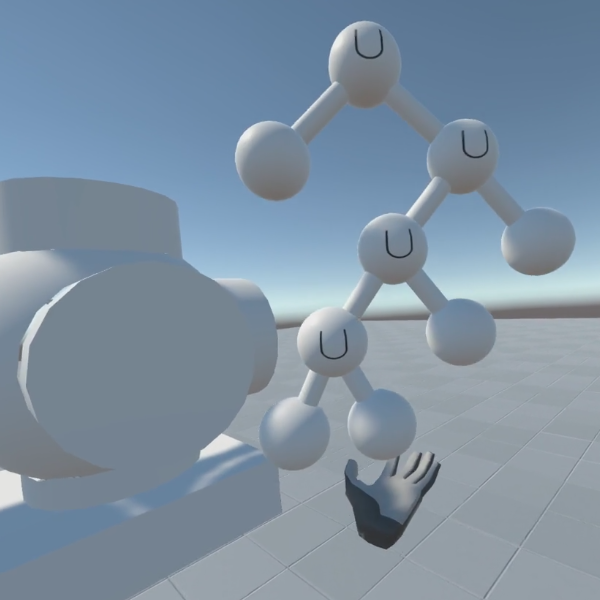}}
	\caption{Activating the the Tree Tool.}
	\label{fig:tree_tool}
\end{figure}

\begin{figure}[h!]
 	\centering
 	\subfloat[Grab Virtual Handle.]{\includegraphics[width=0.45\linewidth]{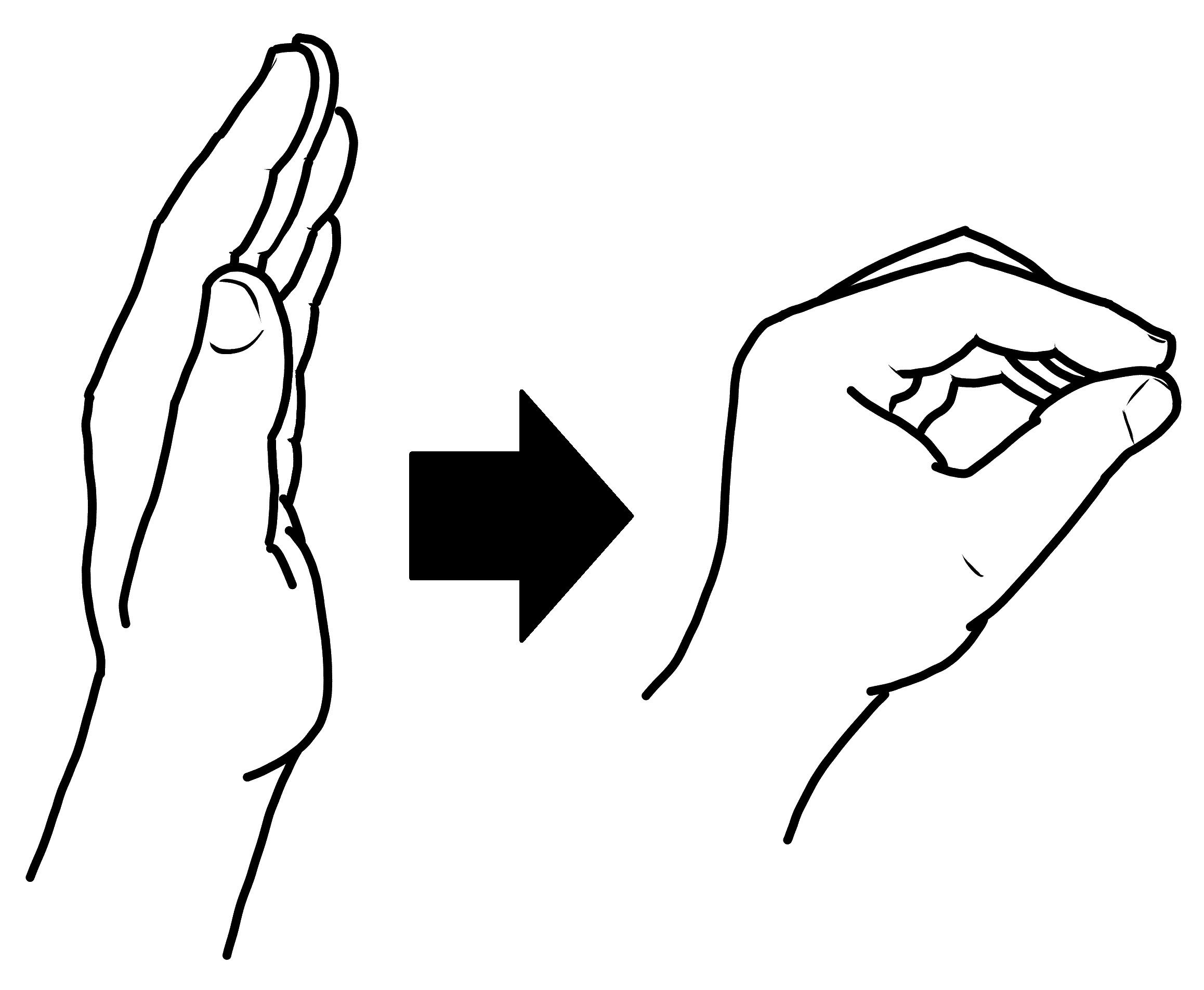}}%
 	\qquad
 	\subfloat[Open Tree Tool.]{\includegraphics[width=0.45\linewidth]{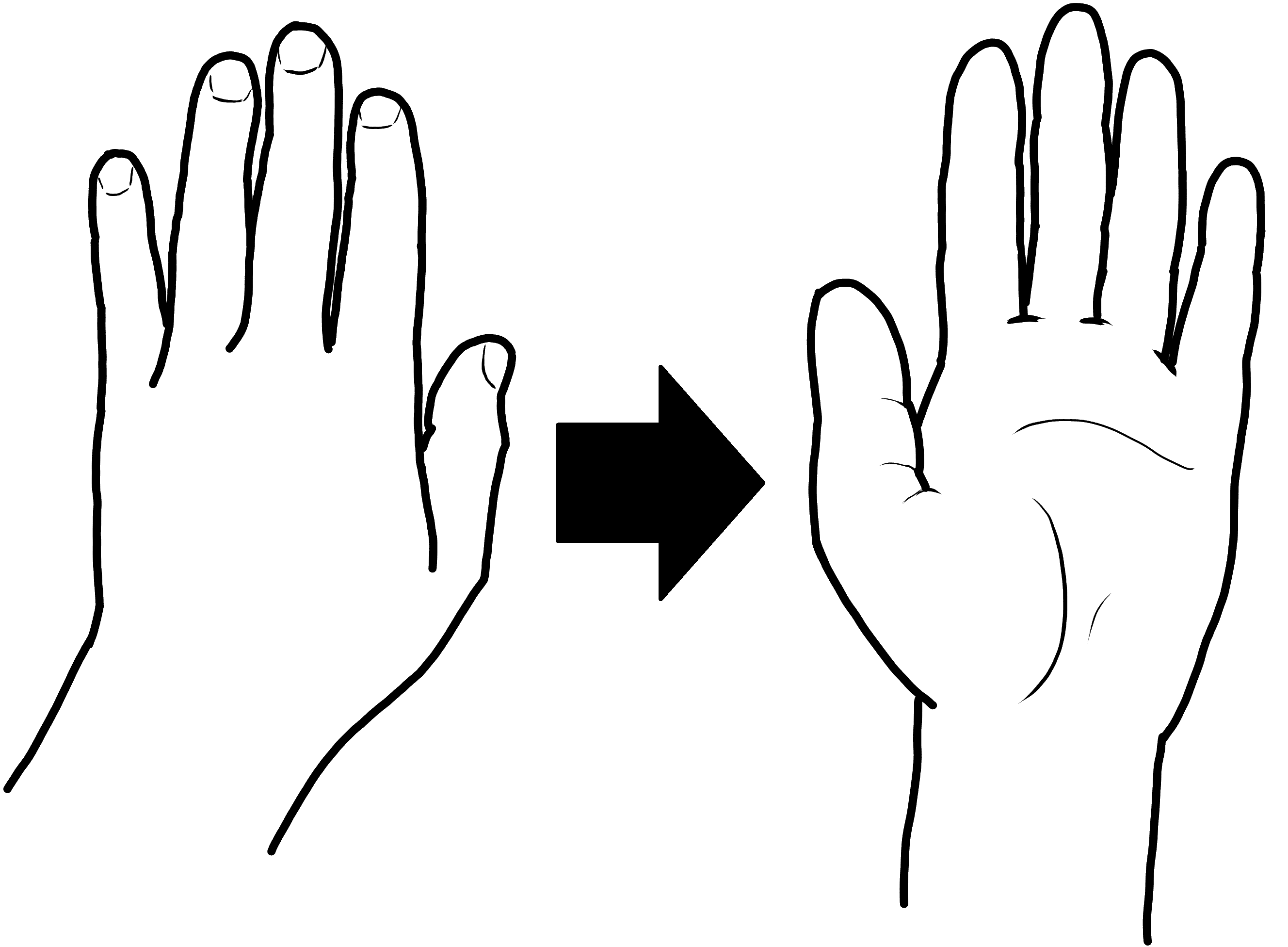}}
   
 	\caption{Used hand gestures.}%
 	\label{fig:gestures}%
 \end{figure}

The tree tool displays a representation of the model's CSG tree using small spheres as nodes. 
It appears above the user's left hand when the user holds it upwards (see Figure \ref{fig:tree_tool} and Figure \ref{fig:gestures} (b)).
Each leaf node corresponds to a primitive, the inner nodes represent the Boolean operators. 
Operation nodes have textures that depict the operation type $(\cup, \cap, -)$.
The user can change the operation type of a node by grabbing the corresponding sphere and invoking one of the following voice commands ('change to union', 'change to inter', 'change to sub').
The tree tool also allows highlighting multiple primitives at once by grabbing their parent node.
Once primitives are highlighted, their corresponding nodes are displayed in green as well (see Figure \ref{fig:tree}). 

\subsubsection{Information Board}
\begin{figure}[h!]
	\centering
	\includegraphics[width=0.43\linewidth]{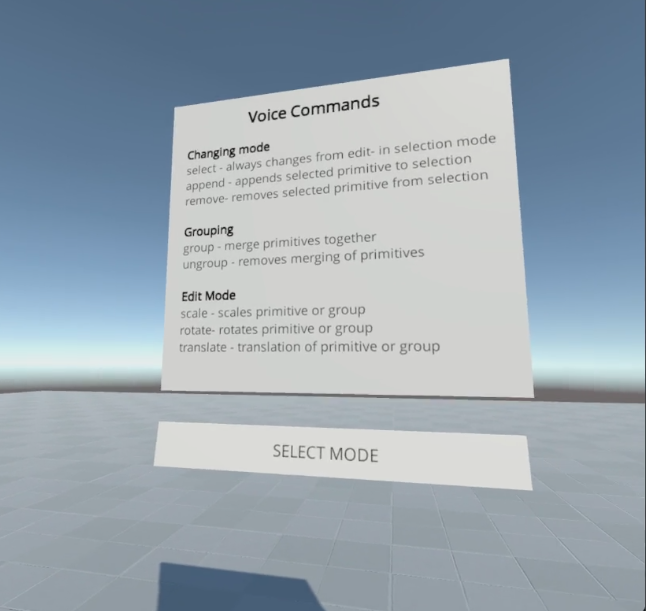}
	\caption{The Information Board as shown to the user.}
	\label{fig:panel}
\end{figure}

The Information Board depicts the current state of the application, the manipulation task, and all voice commands including their explanations (see  Figure \ref{fig:panel}). 
The board is always visible and helps the user to memorize voice commands and to be aware of the current interaction mode.

\section{\uppercase{Evaluation}}\label{sec:evaluation}
A usability study was conducted in order to evaluate the proposed interaction concept. 
Its goal was to validate whether the concept is easy to understand, also for novices in the field of VR and CAD, and whether the combination of hand gesture- and voice control is perceived as a promising idea for intuitive CAD modeling. 
\subsection{Participants}\label{testsetup}
Five student volunteers, three females and two males participated in the study.
Two of them have never used VR headsets before.
Four participants have a background in Computer Science, one participant is a student of Molecular Life Science. 
Details about the participants are depicted in Table \ref{table:participants}.
\begin{table}[h!]
	\centering
	\begin{tabular}{ | l | c | c | c | c | }
		\hline
		\textbf{Gender} &
		\textbf{Age} & 
		\textbf{Background} & 
		\textbf{VR} & 
		\textbf{CAD} \\ [0.2em]
		\hline
		male & 30-40 & UXD & Yes & No \\ [0.2em]
		female & 20-30 & CS & Yes & Yes \\ [0.2em]
		female & 20-30 & MLS & No & No\\ [0.2em]
		female & 20-30 & CS & No & No \\ [0.2em]
		male & 20-30 & CS & No & No \\ [0.2em]
		\hline
	\end{tabular}
	\caption{Overview of the participants. Abbreviations: Background in Computer Science (CS), User Experience Design (UXD) and Molecular Life Science (MLS). Experience in Computer Aided Design (CAD) or Virtual Reality (VR).}
	\label{table:participants}
\end{table}
\subsection{User Study Setup}
\begin{figure}[h!]
	\centering
	\subfloat{\includegraphics[width=0.45\linewidth]{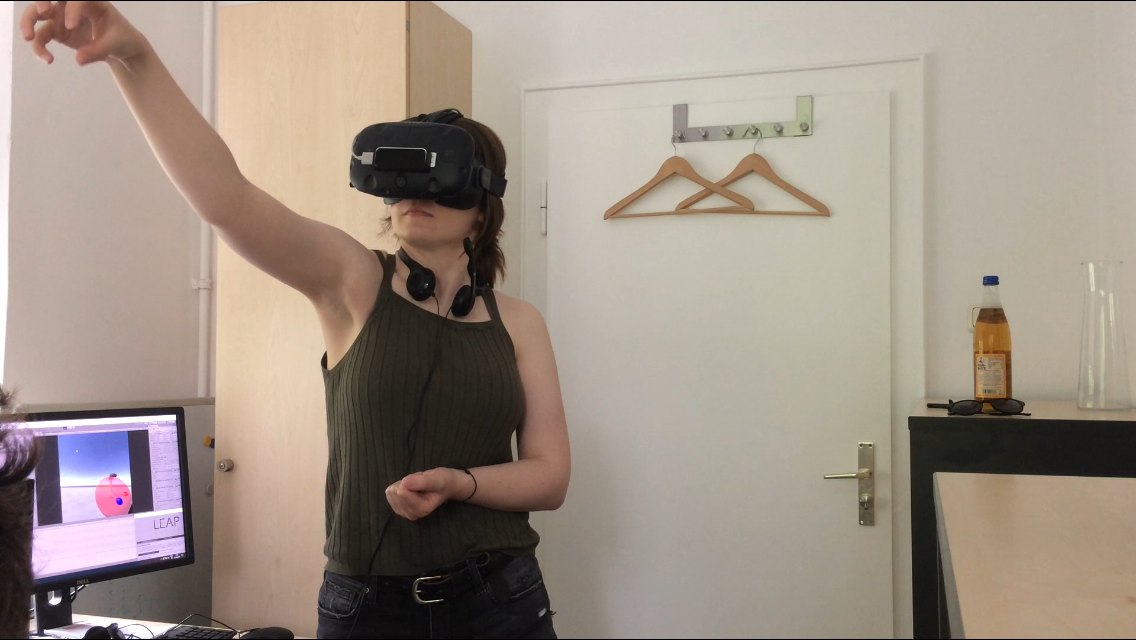}}%
	\qquad
	\subfloat{\includegraphics[width=0.45\linewidth]{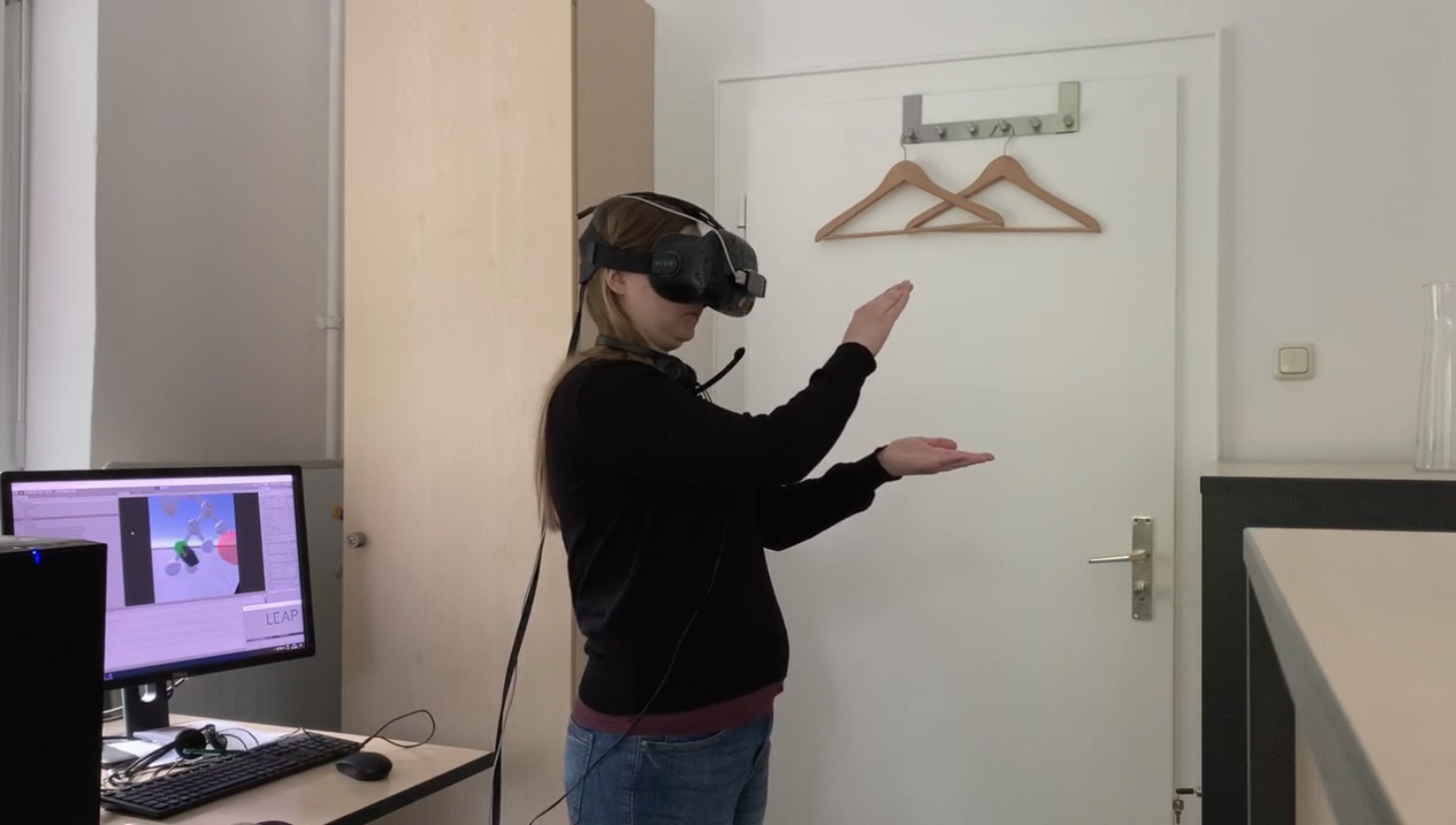}}
	\caption{Participants conducting the user study}%
	\label{fig:participants}%
\end{figure}

\begin{figure}[h!]
	\centering
	\subfloat[Object 1.]{\includegraphics[width=0.25\linewidth]{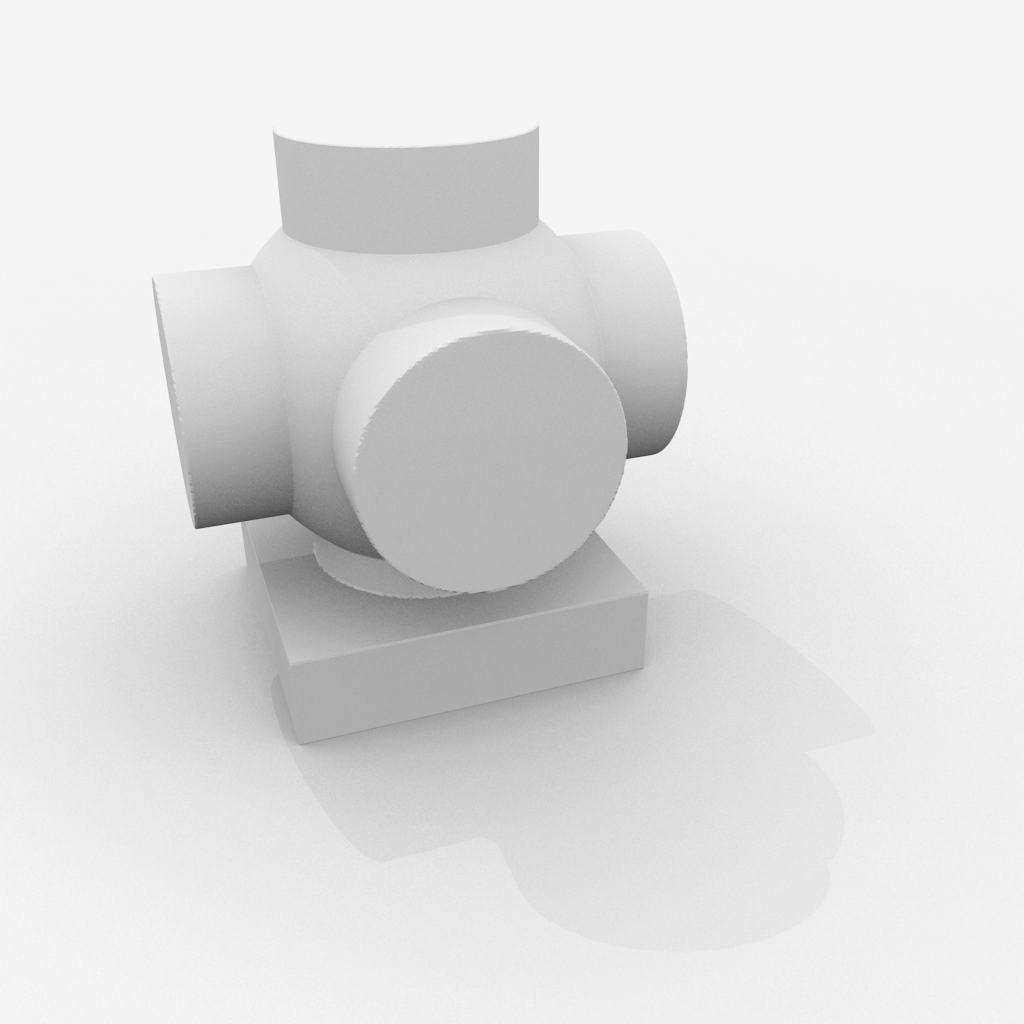}}%
	\qquad
	\subfloat[Object 2.]{\includegraphics[width=0.25\linewidth]{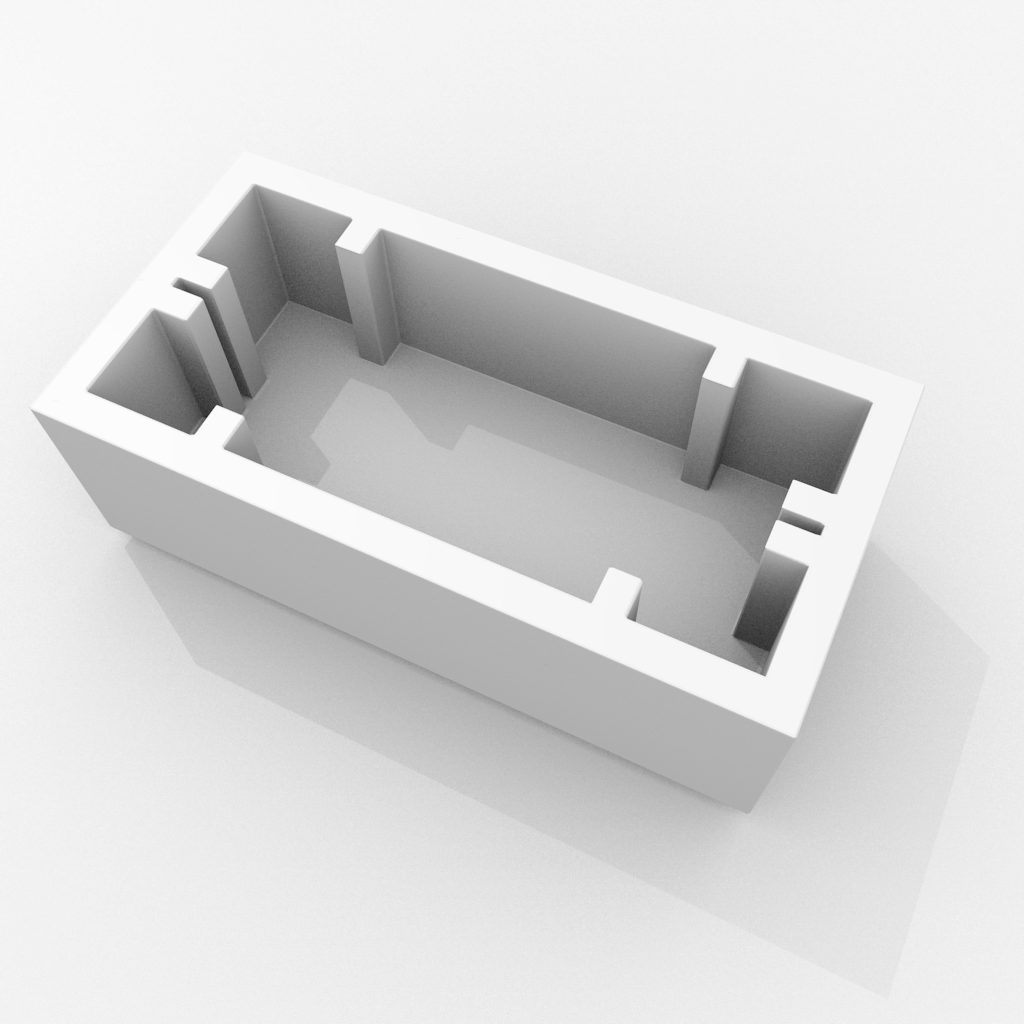}}
	\qquad
	\subfloat[Object 3.]{\includegraphics[width=0.25\linewidth]{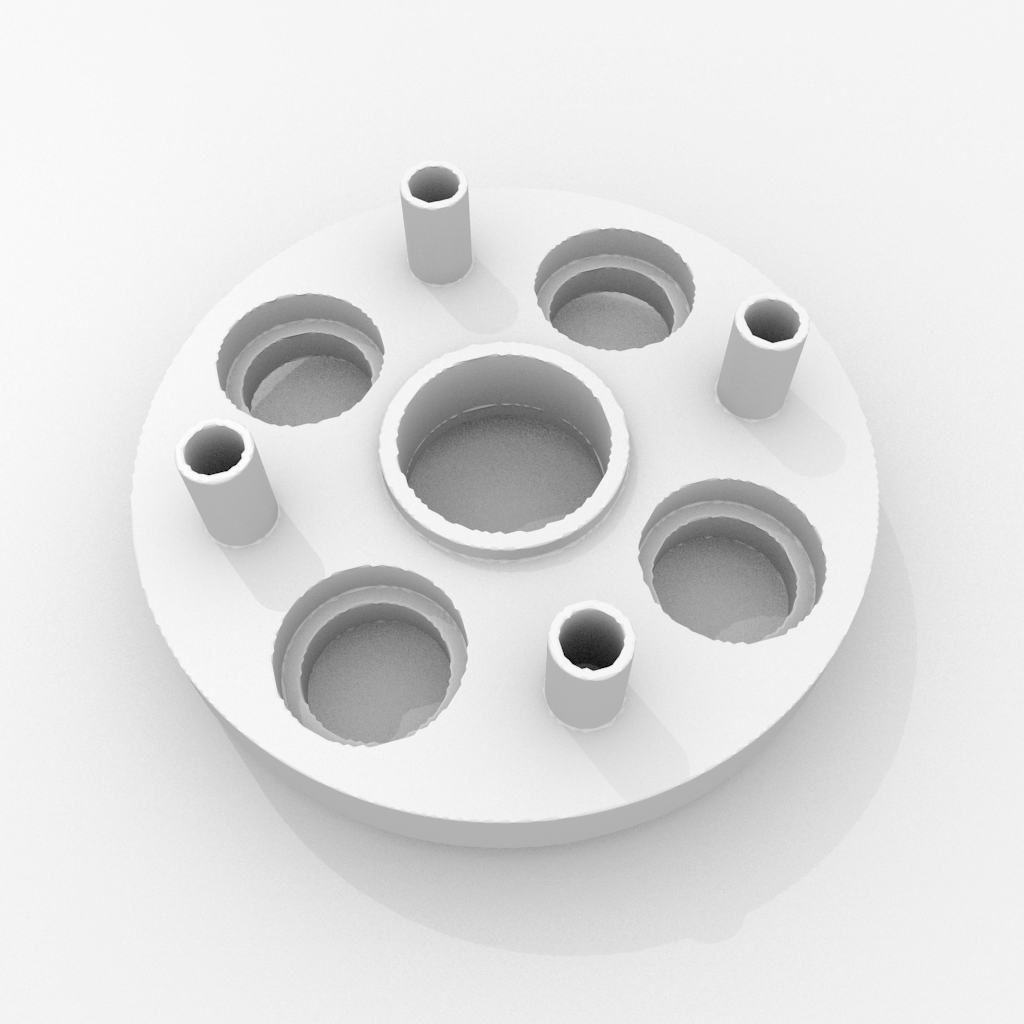}}
	\caption{Objects uses in the user study. Source: \cite{friedrich2019}.}%
	\label{fig:models}%
\end{figure}

Participants could move around freely within a radius of two square meters.
The overall study setting is shown in Figure \ref{fig:participants}.
As an introduction, the participants were shown a simple CAD object (Object 1 in Figure \ref{fig:models}) together with the information board to familiarize themselves with the VR environment. 
Upcoming questions were answered directly by the study director. 
\\
After the introduction, the participants were asked to perform selection and modification tasks (see Table \ref{table:1} for the list of tasks) where they were encouraged to share their thoughts and feelings using a Thinking-Aloud methodology \cite{boren2000}. 
The tasks were picked such that all possible interactions with the prototype were covered.
After completing these tasks, two more objects (Objects 2 and 3 in Figure \ref{fig:models}) were shown to the participants.
The whole process was recorded on video for further research and evaluation purposes. 
\\
The participants' actions, comments and experienced difficulties were logged in order to distill the advantages and disadvantages of the proposed interaction concept. 
In addition, problems that came up during the test could be directly addressed and discussed.
\begin{table}[h!]
	\centering
	\begin{tabular}{ | l |}
		\hline
		 \textbf{Tasks} \\
		  \hline
		Try to select the circle and the cube together. \\ [0.2ex] 
		Deselect the selected primitives. \\  [0.2ex] 
		Select the same primitives using the object tree. \\[0.2ex] 
		Rotate a primitive by 90 degrees. \\[0.2ex] 
		Scale a primitive. \\[0.2ex] 
		Move a cylinder. \\[0.2ex] 
		Change an operator from union to subtraction. \\[0.2ex] 
		\hline
	\end{tabular}
	\caption{Tasks given the participants during user study.}
	\label{table:1}
\end{table}

\subsection{Interviews}\label{interview}
After finishing the aforementioned tasks, an interview was conducted in order to gain further insight into the participants' user experience. 
Each interview started with general questions about pre-existing knowledge of VR and CAD software which should help with the classification of the given answers (see Table \ref{table:questions}).
 
 \begin{table}[h!]
 	\centering
 	\begin{tabular}{ | l |}
 		\hline
 		\textbf{Questions} \\
 		\hline
 		How much experience do you have in virtual \\ reality applications?\\ [0.2em]
 		How much experience do you have in CAD/3D \\ modeling software?\\  [0.2em]
 		How do you see our control method compared to a \\ mouse based interface?\\ [0.2em]
 		What is your opinion on pointer devices?\\[0.2em]
 		\hline
 	\end{tabular}
 	\caption{Selected questions for the interview.}
 	\label{table:questions}
 \end{table}
Statements were assembled based on the catalogue of questions \cite{bossavit2014design} (see Table \ref{table:2}) which served as an opener for discussion.
Interviews were digitally recorded and then transcribed. 

\begin{table}[h!]
	\centering
	\begin{tabular}{ | l |}
	     \hline
		\textbf{Statements} \\
		\hline
		I found the technique easy to understand. \\ [0.2em]
		I found the technique easy to learn. \\   [0.2em]
		I found it easy to select an object. \\ [0.2em]
		I found it easy to group multiple objects. \\ [0.2em]
		I found translation easy to do. \\ [0.2em]
		I found rotation easy to do. \\ [0.2em]
		I found scaling easy to do. \\ [0.2em]
		I found the tree tool easy to understand. \\ [0.2em]
		The tree tool was useful to understand the \\ structure of the object. \\ [0.2em]
		The change of the Boolean operators was easy to \\ understand. \\ [0.2em]
		\hline
	\end{tabular}
	\caption{Statements given to initiate discussion.}
	\label{table:2}
\end{table}

\subsection{Results}\label{results}
In general, the interaction design was well received by the participants even in this early stage of prototype development. 
All users stated that they understood the basic interaction idea and all features of the application. 
Furthermore, they were able to learn the controls quickly.
\\
However, it was observed that users had to get used to the separation between selection and manipulation mode which was not entirely intuitive to them. 
For instance, some participants tried to highlight primitives in the edit mode.  
In this case, additional guidance was needed.
After a short while, however, the user's understanding increased noticeably. 
All users stated that the colorization of objects is an important indication to determine the state of the application and which actions can be executed.
\\
The information board was perceived as being very helpful but was actively used by only three participants.
Two users struggled with the information board  not always being in the field of view which necessitated a turn of the head to see it.
Furthermore, only one user realized that the current state of the application is displayed there as well.
\\
Hand gesture-based model manipulation was very well received by the users. 
The hover and grab gestures did not need any explanation. 
Four users had trouble using their hands in the beginning because they were not within the range of the hand gesture sensor.
\\
The voice control was evaluated positively. 
However, participants reported that the command recognition did not always work.
The overall success rate was $70.7\%$ (see Table \ref{table:4}). 
Interestingly, both male participants have significantly higher success rates than the three female participants which might be due to different pitches of their voices. \\
In addition, some command words like 'append' confused the users, because they expected a new primitive to be appended to the object.
This indicates that an intuitive voice command design is essential but, at the same time, hard to achieve.
\begin{table}[h!]
	\centering
	\begin{tabular}{ | l | c | c | c |}
		\hline
		\textbf{User} & \textbf{recognized} & \textbf{not recog.} &\textbf{Success Rate} \\
		\hline
		P1 & $60$ & $12$ &  $83.3\%$\\ [0.2em]
		P2 & $52$ & $35$ &  $59.8\%$ \\ [0.2em]
		P3 & $49$ & $23$ &  $68.1\%$ \\ [0.2em]
		P4 & $55$ & $27$ &  $67.1\%$ \\ [0.2em]
		P5 & $36$ & $12$ &  $75.0\%$\\ [0.2em]
		\hline
		& $252$ & $109$ & $\varnothing70.7\%$ \\
		\hline
	\end{tabular}
	\caption{Overview of voice control success rates. Success means, that a command was successfully recognized by the system.}
	\label{table:4}
\end{table}
\\
The manipulation technique including the Virtual Handles was also evaluated positively. 
All users highlighted the intuitiveness of translation and scaling in particular. 
Two participants missed the functionality of uniform scaling. 
One participant needed an extra explanation in order to understand the intention of the sphere in the center (axis-free object manipulation). 
The rotation was not fully transparent to the participants. 
All mentioned that they did not immediately recognize the selected rotation axis and instead of rotating the cubes at the end of the handle tried to move the complete handle.
\\
The tree tool was evaluated positively. 
All participants used it to understand and manipulate the structure of the object. 
One participant liked that it was easy-to-use and always accessible. 
Two participants rated it as an intuitive tool without having an in-depth knowledge of what exactly a CSG tree is. 
However, one participant could not directly see the connection between the spheres in the tree and the corresponding primitives.  
Thus, we could observe users hovering random nodes to see the assigned primitive, trying to select specific primitives. 
One participant noted that the tree would be better understood if a small miniature of the primitive was depicted instead of a white sphere. 
On the contrary, three users reported difficulties in understanding the Boolean set-operations.
\\
The participants showed different preferences for highlighting primitives.
Two users rather used the tree tool while the others favored the hover gesture-based approach. 
The selection as well as the de-selection was rated positively mostly for being simple to use.
Especially while dealing with more complex objects, it was perceived more difficult to select the desired primitive. 
This was also due to the size of the elements and due to a smaller distance to the next primitive, which makes hovering more difficult.
It was also observed that in unfavorable cases, the selection hand covered the left hand so that the sensor could no longer detect its palm. 
As a result, the tree disappeared. 
This made it more difficult for users to select a sphere in the tree.

\subsection{Discussion}\label{discussion}

This section discusses the results of the evaluation.

\subsubsection{Limitations}
The accuracy of hand tracking is very important. 
Since only very simple gestures were needed to control the application, the sensor technology showed to be sufficiently robust and precise. 
However, tracking problems occurred repeatedly for no specific or obvious reason. 
This led to problems like unwanted transformations, especially when the end of a gesture was not detected correctly. 
Despite the existence of program and tracking errors, the study participants were able to gain comprehensive insights in the application's interaction flow.
\\
Furthermore, the group of participants was rather small and none of them had any experience with existing mid-air interaction techniques.
Future versions of the prototype should be tested with more users, also taking different target user groups into account (CAD modelling experts and beginners, ...).
In addition, tasks should be performed using standard CAD modeling software in order to be able to compare the approaches properly.

\subsubsection{Findings}
The reported flat learning curve shows the potential of the proposed approach.
Not being dependent on any buttons on physical controllers is advantageous. 
After just a few explanations and a few minutes of training, users were able to modify 3D objects, even those participants without VR or CAD modelling experience.
\\
The Virtual Handle approach is a suitable and intuitive three DOF manipulation tool which can be used for translation, rotation, and scaling in virtual environments. 
For those actions, our prototype needs to be improved, since the implementation of the rotation transformation was not perceived well.
\\
Using additional voice commands for interacting with the system is a good way to simplify the graphical user interface. 
However, voice interfaces must be designed carefully in order to be intuitive while command recognition must work flawlessly.
\\
The examination and manipulation of objects using the CSG tree structure works well even for novices. 
The tree tool can be understood and used without having any knowledge about the theory of CSG trees. 
Especially simple objects can be manipulated very intuitively. 
However, the system quickly reaches its limits if the object consists of many small primitives. 
Furthermore, some users reported that they had difficulties imagining intersection and subtraction operations in advance.
The advantage of the proposed interaction concept becomes apparent when using objects with small corresponding CSG trees. 
The more complex the objects become, the more difficult it is to manipulate them using the proposed method.

\section{\uppercase{Conclusion}}
\label{sec:futurework}
We described and evaluated an interaction design for manipulating CAD models, represented as CSG trees, using hand gestures and voice commands in a VR environment. 
A user study with CAD novices could show that the main concepts come with a flat learning curve and are intuitive to use. 
However, there are some interesting directions for future work:
\\
An important additional function is the ability to append/remove primitives to/from the CSG tree using specific voice commands - also for determining shape and position.
This would also necessitate a solution for the large model problem for both, interacting with the model directly as well as with the CSG tree, which could be solved by adapting classical zoom \& scroll paradigms to VR environments (complementary to the Go-Go and Ray-casting techniques described in Section \ref{sec:related_work}). 
Furthermore, our manipulation technique has not yet solved the general precision problem: 
Exact rotations, for example, are very difficult to realize. 
One idea would be to use a 'magnetic' grid, which is placed at the most important surface points of the object. As soon as transformations near these points are completed, the primitive can snap automatically to these points.
A system based on Natural Language Processing could analyze and interpret not only simple commands but complete sentences.
Users would not have to stick to exact voice commands anymore but would rather describe which action to perform. 
The application gives specific feedback by executing the action or giving further instructions on how to execute it.
This could also help solving the aforementioned precision problem since the user could directly specify coordinates and rotation angles by saying the precise numbers. 

\section*{\uppercase{Acknowledgements}}
\noindent We thank [Anonymous] for creating the hand gesture illustrations in Figure \ref{fig:gestures}.

\bibliographystyle{apalike}
{\small
\bibliography{bib}}

%

\end{document}